\documentclass[11pt]{article}

\usepackage[top = 30 mm, bottom = 30mm, left = 20mm, right=20mm]{geometry}
\usepackage[utf8]{inputenc}
\usepackage[english]{babel}
\usepackage{lmodern}
\usepackage{ragged2e}

\usepackage{cite}
\usepackage{multirow}

\usepackage{graphicx}
\usepackage{xcolor}
\usepackage{float}
\usepackage{color}

\definecolor{orange}{rgb}{0.9,0.2,0}
\definecolor{brown}{rgb}{0.7,0.3,0.2}
\definecolor{fuxia}{rgb}{1,0,1}
\definecolor{skyblue}{rgb}{0,0.1,0.9}
\definecolor{violetred}{rgb}{0.8,0.13,0.56}
\definecolor{deeppink}{rgb}{1.00,0.08,0.5}
\definecolor{pink}{rgb}{1.00,0.75,0.80}
\definecolor{orchid}{rgb}{0.85,0.44,0.84}
\definecolor{lightpink}{rgb}{1.00,0.71,0.76}
\definecolor{bluish}{rgb}{0,0.6,0.8}
\usepackage{charter}

\usepackage{amsmath, amssymb}
\usepackage{bm}

\usepackage{jheppub}
\usepackage{slashed}
\usepackage{latexsym}

\numberwithin{equation}{section}

\title{Addendum to ``Ultraviolet behaviour of Higgs inflation models"}
\author[a, b]{Ignatios Antoniadis}
\author[a]{Anthony Guillen}
\author[c, d]{Kyriakos Tamvakis}

\affiliation[a]{Laboratoire de Physique Th\'eorique et Hautes Energies - LPTHE\\
Sorbonne Universit\'e, CNRS, 4 Place Jussieu, 75005 Paris, France}
\affiliation[b]{Nordita,
Stockholm University and KTH Royal Institute of Technology\\
Hannes Alfv\'ens v\"ag 12, 106 91 Stockholm, Sweden}
\affiliation[c]{Physics Department, University of Ioannina, 45110, Ioannina, Greece}
\affiliation[d]{CERN,Theoretical Physics Department, 1211 Geneva 23, Switzerland}

\emailAdd{antoniad@lpthe.jussieu.fr}
\emailAdd{anthony.guillen@protonmail.com}
\emailAdd{tamvakis@uoi.gr}

\date{} 

\abstract{This article is an addendum to \cite{Antoniadis}. We extend our computation of the on-shell scattering amplitudes in an arbitrary inflaton background $\bar{\phi}_1$. Although the effective Einstein frame cutoff for $\bar{\phi}_1>>M_P/\sqrt{\xi}$ turns out to be $\bar{\phi}_1$ or $\xi\bar{\phi}_1$ for the $U(1)$ model, this is not the case for the realistic doublet Higgs model where the effective Einstein frame cutoff turns out to be the standard $M_P/\sqrt{\xi}$ for both the Palatini and metric formulations. However, as it has been pointed out in \cite{Antoniadis} the background $\bar{\phi}_1$ is the effective Jordan frame cutoff in all cases for both the Palatini and metric formalisms.}

\begin{document}

\flushright NORDITA 2022-016

\justifying

\maketitle

\vfill\section{Framework}

 We start by reminding the set-up of our computation. We consider the usual action of Higgs inflation, ignoring the potential to focus on the effects of the non minimal coupling
\begin{equation}
\mathcal{S} = \int d^4x \sqrt{-g}\left(\frac{1}{2}M_P^2\Omega^2 R - |\partial H|^2\right) \hspace{\baselineskip}\text{with}\hspace{\baselineskip}\Omega^2 = 1 + \frac{2\xi|H|^2}{M_P^2}
\end{equation}

Calculations with this action are simpler in the Einstein frame to which we go with the transformation:
\begin{equation}\label{conformal_transformation}
g_{\mu\nu} \rightarrow \Omega^{-2}g_{\mu\nu}, \hspace{\baselineskip} g^{\mu\nu} \rightarrow \Omega^{2}g^{\mu\nu}, \hspace{\baselineskip} \sqrt{-g}\rightarrow \Omega^{-4}\sqrt{-g}, \hspace{\baselineskip}R \rightarrow \Omega^2 R - \underline{3/2\Omega^{-2}(\partial\Omega^2)^2}
\end{equation}

The underlined term in the transformation of $R$ being present in the metric formalism but not in Palatini. From now, we will only work in the Einstein frame. The action then becomes
\begin{equation}\label{Einstein_frame_action}
\mathcal{S} = \int d^4x \sqrt{-g}\left(\frac{1}{2}M_P^2 R - \frac{1}{\Omega^2}|\partial H|^2 - \underline{\frac{3M_P^2}{4}\frac{(\partial\Omega^2)^2}{\Omega^4}}\right) 
\end{equation}

In \cite{Antoniadis} we considered the Higgs to be a complex singlet $H = (\phi_1 + i\phi_2)/\sqrt{2}$. Although not "realistic", it seemed enough to make our point on the ultraviolet behaviour. Then, we introduced by hand a large, inflationary background $2\bar H^2 \gg M_P^2/\xi$, taken in the direction of $\phi_1$ without loss of generality to be $\phi_1 = \bar\phi_1 + \phi_1'$. In this notation $\phi_1'$ is then the physical Higgs and $\phi_2$ is a Goldstone boson. After doing so, we expanded the different terms of the action to find the following interactions between $\phi_1'$ and $\phi_2$, using $\bar\phi_1^2 \gg M_P^2/\xi$. This results into several simplifications. For instance, in the Palatini formalism we obtain
the following lagrangian terms in a self-explanatory notation:
\begin{equation}\label{interactions_previous_computation}
\mathcal{L}_{\chi_1'^3} = \frac{\sqrt{\xi}}{M_P}\chi_1'(\partial\chi_1')^2, \hspace{\baselineskip}
\mathcal{L}_{\chi_1'\chi_2^2} = \frac{\sqrt{\xi}}{M_P}\chi_1'(\partial\chi_2)^2, \hspace{\baselineskip}
\mathcal{L}_{\chi_1'^2\chi_2^2} = -\frac{3\xi}{2M_P^2}\chi_1'^2(\partial\chi_2)^2 + \frac{\xi}{2M_P^2}\chi_2^2(\partial\chi_1)^2
\end{equation}

Here, $\chi_1' = M_P\phi_1'/(\sqrt{\xi}\bar\phi_1)$ and $\chi_2 = M_P\phi_2/(\sqrt{\xi}\bar\phi_1)$ are the canonically normalized counterparts of $\phi_1', \phi_2$. From these interactions we deduce the relevant vertices. For instance, the vertex coming from $\mathcal{L}_{\chi_1'^3}$ is
\begin{equation}
V_{\chi_1'^3} = -\frac{2i\sqrt{\xi}}{M_P}\left(p_{\chi_1', 1}\cdot p_{\chi_1', 2} + p_{\chi_1', 1}\cdot p_{\chi_1', 3} + p_{\chi_1', 2}\cdot p_{\chi_1', 3}\right)\,,
\end{equation}

where the $p_{\chi_1', i}$ are the momenta entering the vertex. Here, the minus sign is a $i^2$ from the two derivatives. \\

Using the above vertices and the propagator $G_{\chi_i}(p) = -{i}/{p^2}$, we proceed to compute the $\chi_1'\chi_2 \rightarrow \chi_1'\chi_2$ amplitude (the $\chi_1'\chi_1' \rightarrow \chi_1'\chi_1'$ vanishes due to crossing symmetry, and the same for $\chi_2$). Ignoring graviton exchange that does not depend on $\xi$ in the Einstein frame, there are four graphs corresponding to this amplitude at tree level: one for the quartic vertex, one for $\chi_1'$ exchange in the $t$ channel, two for $\chi_2$ exchange in the $s$ and $u$ channels. We denote the corresponding subamplitudes $\mathcal{M}^4, \mathcal{M}^t_{\chi_1'}, \mathcal{M}^s_{\chi_2}, \mathcal{M}_{\chi_2}^u$. Computing them yields:
\begin{equation}\label{subamplitudes_previous_computation}
\mathcal{M}^4 = -\frac{2\xi t}{M_P^2}, \hspace{\baselineskip} \mathcal{M}^t_{\chi_1'} = \frac{\xi t}{M_P^2}, \hspace{\baselineskip} \mathcal{M}^s_{\chi_2} = -\frac{\xi s}{M_P^2}, \hspace{\baselineskip} \mathcal{M}^u_{\chi_2} = -\frac{\xi u}{M_P^2}
\end{equation}

here, $s = -(p_1+p_2)^2, t = -(p_1-p_3)^2, u=-(p_1-p_4)^2$ are the Mandelstam variables in mostly plus convention (in \cite{Antoniadis}, we used a different definition, without the minus sign, but this does not make much difference), where $p_1, p_2$ (resp. $p_3, p_4$) are the momenta of the incoming (resp. outgoing) $\chi_1',\chi_2$. Correcting a sign error in the scalar propagator made in \cite{Antoniadis}, we obtain a vanishing amplitude
\begin{equation}
\mathcal{M}(\chi_1'\chi_2 \rightarrow \chi_1'\chi_2) = \mathcal{M}^4 + \mathcal{M}^t_{\chi_1'} + \mathcal{M}^s_{\chi_2} + \mathcal{M}_{\chi_2}^u = 0
\end{equation}

This computation was made in the Palatini formulation but the same cancellation occurs in the metric formalism. As a consequence of this, demanding tree-level unitarity, it seems that the usual cutoff $\Lambda\sim M_P/\sqrt{\xi}$ of Higgs inflation in a large background should be lifted up to $M_P$, where graviton interactions start to cause trouble. But before claiming this, we should make more precise the behavior in the $\bar\phi_1 \gg M_P/\sqrt{\xi}$ limit, and check what happens if we consider the Higgs to be a complex doublet, as it is in the Standard Model. This is what the next two sections are devoted to.
\section{Extension to general background}

Let us now extend the previous calculation to a general background $\bar\phi_1$, that is, without making the assumption $\bar\phi_1^2 \gg M_P^2/\xi$. The only requirement is that this background varies slowly in time and space, for it to be considered constant in the region and timescale of the interactions between $\phi_1'$ and $\phi_2$. Setting
\begin{equation}
\bar\Omega^2 = 1 + \frac{\xi\bar\phi_1^2}{M_P^2} \hspace{\baselineskip}\text{and}\hspace{\baselineskip} x^2 = \frac{\xi \bar\phi_1^2}{M_P^2\bar\Omega^2}\,,
\end{equation}
we proceed to the computation just as before first in the Palatini and then in the metric formalism.

\subsection{Palatini formalism}

In the Palatini formalism, after developing (\ref{Einstein_frame_action}), we extract the relevant interactions of the canonically normalized fields for the process $\chi_1'\chi_2 \rightarrow \chi_1'\chi_2$ 
\begin{equation}\label{interactions_Palatini_background}
\mathcal{L}_{\chi_1'^3} = \frac{x\sqrt{\xi}}{M_P}\chi_1'(\partial\chi_1')^2, \hspace{\baselineskip}
\mathcal{L}_{\chi_1'\chi_2^2} = \frac{x\sqrt{\xi}}{M_P}\chi_1'(\partial\chi_2)^2, \hspace{\baselineskip}
\mathcal{L}_{\chi_1'^2\chi_2^2} = \frac{(1-4x^2)\xi}{2M_P^2}\chi_1'^2(\partial\chi_2)^2 + \frac{\xi}{2M_P^2}\chi_2^2(\partial\chi_1)^2
\end{equation}

When $\bar\phi_1^2 \rightarrow M_P^2/\xi$ we have $x \rightarrow 1$ and of course we recover (\ref{interactions_previous_computation}). Here as before, the $\chi_i = \phi_i/\bar\Omega$ are the canonically normalized counterparts of the $\phi_i$. From these interactions, we deduce the corresponding vertices and using the same notation as in equation (\ref{subamplitudes_previous_computation}), we compute the following subamplitudes
\begin{equation}
\mathcal{M}^4 = \frac{2(1-2x^2)\xi t}{M_P^2}, \hspace{\baselineskip}
\mathcal{M}^t_{\chi_1'} = \frac{\xi x^2 t}{M_P^2}, \hspace{\baselineskip}
\mathcal{M}^s_{\chi_2} = -\frac{\xi x^2 s}{M_P^2}, \hspace{\baselineskip}
\mathcal{M}^u_{\chi_2} = -\frac{\xi x^2 u}{M_P^2}
\end{equation}

Adding them together yields:
\begin{equation}
\mathcal{M}(\chi_1'\chi_2 \rightarrow \chi_1'\chi_2) = \frac{2(1-x^2)\xi t}{M_P^2}\,,
\end{equation}

which is vanishing when $x \rightarrow 1$. More precisely, we can develop $x^2 \sim 1 - M_P^2/(\xi\bar\phi_1^2)$ when $\bar\phi_1^2 \gg M_P^2/\xi$, so that:
\begin{equation}\label{amplitude_Palatini_background}
\mathcal{M}(\chi_1'\chi_2 \rightarrow \chi_1'\chi_2) \simeq \frac{2t}{\bar\phi_1^2} \sim \frac{E^2}{\bar\phi_1^2}\hspace{\baselineskip}\mathrm{when}\hspace{\baselineskip}\bar\phi_1^2 \gg M_P^2/\xi
\end{equation}

It then turns out that in this limit, the energy at which $\mathcal{M} \sim 1$, identified with the cutoff of the effective theory, is $\Lambda \sim \bar\phi_1$ in the Einstein frame. This is above the usual $M_P/\sqrt{\xi}$. We will discuss it more later.

\subsection{Metric formalism}

In the metric formalism, after developing (\ref{Einstein_frame_action}), the relevant interactions for $\chi_1'\chi_2 \rightarrow \chi_1'\chi_2$ are:
\begin{eqnarray}
\mathcal{L}_{\chi_1'^3} &=& \frac{(1-6\xi+12\xi x^2)x\sqrt{\xi}}{(1+6\xi x^2)^{3/2}M_P}\chi_1'(\partial\chi_1')^2\\
\mathcal{L}_{\chi_1'\chi_2^2} &=& \frac{x\sqrt{\xi}}{(1+6\xi x^2)^{1/2}M_P}\chi_1'(\partial\chi_2)^2 
- \frac{6\xi^{3/2}x}{(1+6\xi x^2)^{1/2}M_P}\chi_2(\partial\chi_1'\cdot\partial\chi_2)\label{interactions_metric_background}
\end{eqnarray}

and
\begin{equation}
\mathcal{L}_{\chi_1'^2\chi_2^2} =
\frac{(1-4x^2)\xi}{2(1+6\xi x^2)M_P^2}\chi_1'^2(\partial\chi_2)^2 
+ \frac{(1+12\xi x^2)\xi}{2(1+6\xi x^2)M_P^2}\chi_2^2(\partial\chi_1')^2
- \frac{6(1 - 4x^2)\xi^2}{(1+6\xi x^2)M_P^2}\chi_1'\chi_2(\partial\chi_1'\cdot\partial\chi_2)\,,
\end{equation}

where $\chi_1' = (1+6\xi x^2)\phi_1'/\bar\Omega$ and $\chi_2 = \phi_2/\bar\Omega$ are the canonically normalized counterparts of the $\phi'_i$ and $\phi_2$, respectively.
From these interaction terms, using the same notation as in equation (\ref{subamplitudes_previous_computation}), we compute the following subamplitudes
\begin{eqnarray}
&&\mathcal{M}^4 = \frac{2(1-2x^2)(1+3\xi)}{1+6\xi x^2}\frac{\xi t}{M_P^2}, \hspace{\baselineskip}
\mathcal{M}^t_{\chi_1'} = \frac{(1+6\xi)(1-6\xi+12\xi x^2)}{(1+6\xi x^2)^2}\frac{\xi x^2 t}{M_P^2}\\
&&\mathcal{M}^s_{\chi_2} = -\frac{\xi x^2 s}{(1+6\xi x^2)M_P^2}, \hspace{\baselineskip}
\mathcal{M}^u_{\chi_2} = -\frac{\xi x^2 u}{(1+6\xi x^2)M_P^2}
\end{eqnarray}
Summing them yields
\begin{equation}\label{amplitude_metric_background}
\mathcal{M}(\chi_1'\chi_2 \rightarrow \chi_1'\chi_2) = \frac{2((1-x^2)+3\xi(1-x^4))\xi t}{(1+6\xi x^2)^2M_P^2}\,.
\end{equation}

A first observation here is that the terms proportional to $\xi^3$ cancel in the numerator, meaning that for fixed $x$, in the limit where $\xi \gg 1$, we have $\mathcal{M}\sim E^2/M_P^2$, giving a cutoff $\Lambda \sim M_P$, above $M_P/\sqrt{\xi}$. Then, $\mathcal{M}$ vanishes when $x \rightarrow 1$. More precisely, developing $x^2 \sim 1 - M_P^2/(\xi\bar\phi_1^2)$, when $\bar\phi_1^2 \gg M_P^2/\xi$, we obtain
\begin{equation}\label{amplitude_metric_background}
\mathcal{M}(\chi_1'\chi_2 \rightarrow \chi_1'\chi_2) \simeq \frac{2t}{(1+6\xi)\bar\phi_1^2} \sim \frac{E^2}{\xi\bar\phi_1^2}\hspace{\baselineskip}\mathrm{when}\hspace{\baselineskip}\bar\phi_1^2 \gg M_P^2/\xi 
\hspace{\baselineskip}\mathrm{and}\hspace{\baselineskip}\xi \gg 1\,\,.
\end{equation}
Then it turns out that in this limit the Einstein frame cutoff of the effective theory is $\Lambda \sim \bar\phi_1\sqrt{\xi}$. Thus, in both the Palatini and metric formalisms, cancellations in the amplitude of $\chi_1'\chi_2 \rightarrow \chi_1'\chi_2$ scattering between the canonically normalized physical Higgs and Goldstone boson suggest a cutoff scale $\Lambda\sim\bar\phi_1$ or $\lambda\sim\bar\phi_1\sqrt\xi$ that can be much higher than the usual $M_P/\sqrt{\xi}$ in a large background. We will now check whether this result holds if we consider the scalar to be an $SU(2)$ doublet, as is the Higgs in the Standard Model.

\section{Extension to a complex Higgs doublet}

When $H$ is a complex $SU(2)$ doublet, we can parameterize it as
\begin{equation}
H = \frac{1}{\sqrt{2}}\left(\begin{matrix}\phi_1 + i\phi_2\\\phi_3 + i\phi_4\end{matrix}\right)
\end{equation}

Then, we may, without loss of generality, introduce a background in the direction of $\phi_1$ as before. In this case, the Goldstone bosons $\phi_2, \phi_3, \phi_4$ play exactly the same role and are interchangeable, so that the amplitudes of e.g. $\chi_1'\chi_3 \rightarrow \chi_1'\chi_3$ are the same as in equations (\ref{amplitude_Palatini_background}) and (\ref{amplitude_metric_background}). Note however that now there are new possible scattering processes between different Goldstone bosons. Without loss of generality we shall consider $\chi_2\chi_3 \rightarrow \chi_2\chi_3$ .

\subsection{Palatini formalism}

In this case, the relevant interactions we obtain after developing the action (\ref{Einstein_frame_action}) are $\mathcal{L}_{\chi_1'\chi_3^2}$, which is the same as $\mathcal{L}_{\chi_1'\chi_2^2}$ in (\ref{interactions_Palatini_background}), and
\begin{equation}
\mathcal{L}_{\chi_2^2\chi_3^2} = \frac{\xi}{2}\chi_2^2(\partial\chi_3)^2 + \frac{\xi}{2}\chi_3^2(\partial\chi_2)^2\,.
\end{equation}

The corresponding vertices follow as previously. Ignoring graviton exchange as before, there are now two graphs corresponding to the $\chi_2\chi_3 \rightarrow \chi_2\chi_3$ amplitude at tree level: one for the quartic vertex and one for $\chi_1'$ exchange in the $t$ channel. We denote the corresponding subamplitudes $\mathcal{M}^4$ and $\mathcal{M}_{\chi_1'}^t$ respectively. We get
\begin{equation}\label{amplitude_Palatini_doublet}
\mathcal{M}^4 = \frac{2\xi t}{M_P^2} \hspace{\baselineskip}\text{and}\hspace{\baselineskip}
\mathcal{M}_{\chi_1'}^t = -\frac{\xi x^2t}{M_P^2}\hspace{\baselineskip}\text{so that}\hspace{\baselineskip}
\mathcal{M}(\chi_2\chi_3 \rightarrow \chi_2\chi_3) = \frac{(2-x^2)\xi t}{M_P^2}
\end{equation}
This time, the amplitude does not vanish when $x \rightarrow 1$, so $\mathcal{M}\sim \xi E^2/M_P^2$, giving an Einstein frame cutoff at $\Lambda\sim M_P/\sqrt{\xi}$. 

\subsection{Metric formalism}
In this case, the relevant interactions we obtain after developing the action (\ref{Einstein_frame_action}) are $\mathcal{L}_{\chi_1'\chi_3^2}$, which is the same as $\mathcal{L}_{\chi_1'\chi_2^2}$ in (\ref{interactions_metric_background}), and
\begin{equation}
\mathcal{L}_{\chi_2^2\chi_3^2} = \frac{\xi}{2}\chi_2^2(\partial\chi_3)^2 + \frac{\xi}{2}\chi_3^2(\partial\chi_2)^2 - 6\xi^2\chi_2\chi_3(\partial\chi_2\cdot\partial\chi_3)
\end{equation}
Using the same notations as in (\ref{amplitude_Palatini_doublet}), we compute the following subamplitudes
\begin{equation}
\mathcal{M}^4 = \frac{2(1+3\xi)\xi t}{M_P^2} \hspace{\baselineskip}\text{and}\hspace{\baselineskip}
\mathcal{M}_{\chi_1'}^t = -\frac{(1+6\xi)^2\xi x^2 t}{(1+6\xi x^2)M_P^2}\,.
\end{equation}
Summing them yields
\begin{equation}\label{amplitude_metric_doublet}
\mathcal{M}(\chi_2\chi_3\rightarrow\chi_2\chi_3) = \frac{2-x^2 + 6\xi}{1+6\xi x^2}\frac{\xi t}{M_P^2}\,.
\end{equation}
Again, this does not vanish in the large background limit $x \rightarrow 1$ but goes to $\mathcal{M}\sim\xi t/M_P^2$. Therefore, in this limit or in the limit where $\xi \gg 1$, we have $\mathcal{M}\sim \xi E^2/M_P^2$, giving the usual cutoff at $\Lambda\sim M_P/\sqrt{\xi}$. 

\section{Discussion and conclusion}

Let us collect the results (\ref{amplitude_Palatini_background}), (\ref{amplitude_metric_background}), (\ref{amplitude_Palatini_doublet}), (\ref{amplitude_metric_doublet}) at leading order when $\xi \gg 1$ and $\bar\phi_1^2\gg M_P^2/\xi$:

\begin{center}\begin{tabular}{c||c|c||c|c||c|c}
\multirow{2}{1pt}{}& \multicolumn{2}{c||}{amplitude $\mathcal{M}$} & \multicolumn{2}{c||}{cutoff $\Lambda$ (Einstein)} & \multicolumn{2}{c}{cutoff $\Lambda$ (Jordan)}\\
& Palatini & metric & Palatini & metric & Palatini & metric \\
\hline\hline
& & & & & &\\
$\chi_1'\chi_2 \rightarrow \chi_1'\chi_2$ & $E^2/\bar\phi_1^2$ & $E^2/(\xi\bar\phi_1^2)$ & $\bar\phi_1$ & $\bar\phi_1\sqrt{\xi}$ & $\bar\phi_1^2\sqrt{\xi}/M_P$ & $\xi\bar\phi_1^2/M_P$\\
& & & & & &\\\hline& & & & & &\\
$\chi_2\chi_3 \rightarrow \chi_2\chi_3$ & $\xi E^2/M_P^2$ & $\xi E^2/M_P^2$ & $M_P/\sqrt{\xi}$ & $M_P/\sqrt{\xi}$ & $\bar\phi_1$ & $\bar\phi_1$\\
& & & & & &
\end{tabular}\end{center}

Here, in addition to the amplitudes and associated cutoff in the Einstein frame, in which we worked here, we reported cutoff in the Jordan frame. The two are simply linked by the conformal transformation (\ref{conformal_transformation}):
\begin{equation}
\Lambda^{(J)} = \bar\Omega\Lambda^{(E)} \hspace{\baselineskip}\text{and}\hspace{\baselineskip} \bar\Omega\simeq \frac{\bar\phi_1\sqrt{\xi}}{M_P} \hspace{\baselineskip}\text{when}\hspace{\baselineskip}\bar\phi_1^2 \gg \frac{M_P^2}{\xi}\,.
\end{equation}
This can help to compare with \cite{Antoniadis}, where as it has been pointed out the background $\bar{\phi}_1$ is the effective Jordan frame cutoff in all cases for both the Palatini and metric formalisms. However, summarising, we stick to the Einstein frame. 
In \cite{Antoniadis} we investigated a simplified model where the Higgs was a singlet. In this case, only the first line of the table is relevant. We find that the cutoff of this simplified model can be much higher, at $\bar\phi_1$ (Palatini) or $\bar\phi_1\sqrt{\xi}$ (metric). When the Higgs is, rightfully, considered to be a doublet, there are new nonvanishing Goldstone-Goldstone amplitudes (second line of the table) for which no cancellation occurs yielding the usual cutoff at $M_P/\sqrt{\xi}$ as in \cite{Ito}. Therefore, the singlet model has a softer behaviour than the more realistic doublet model at high energies.

\section*{Acknowledgments}

We thank Asuka Ito, Wafaa Khater and Syksy Räsänen for communications allowing us to find a sign error.

\bibliographystyle{unsrt}
\bibliography{bibliographie}

\vfill

\end{document}